\documentclass[amsmath,amssymb,epsfig,aps,pra,twocolumn]{revtex4-2}
\usepackage{ifpdf}
\usepackage{graphicx}
\usepackage{amssymb}
\usepackage{amsmath}
\usepackage{color}
\usepackage{xcolor}
\usepackage{multirow}
\usepackage{afterpage}
\usepackage{hyperref}
\usepackage{ulem}
\usepackage{cancel}
\begin{document}
\title[STABILITY DOMAIN OF 3D BEC]{The effects of elastic and inelastic collisions in two- and three-body interactions on the stability of 3D Bose-Einstein condensates}
\author{R. Sasireka$^1$, O.T. Lekeufack,$^2$, S. Sabari$^3$, and A. Uthayakumar$^1$}
\affiliation{
$^{1}$Department of Physics, Presidency College (Autonomous), Chennai - 600005, India \\ 
$^{2}$Department of Physics, Faculty of Science, University of Douala, P.O. Box 24157 Douala, Cameroon \\
$^{3}$Instituto de F\'\i sica Te\'{o}rica, UNESP -- Universidade Estadual Paulista, 01140-070 S\~{a}o Paulo, Brazil
}
\date{\today}

\begin{abstract}
In this paper, we study the stability of three-dimensional Bose-Einstein condensates of finite temperatures at which both elastic and inelastic collisions are taken into account. The modeled governing Gross-Pitaevski equation reveals inclusion of both real and imaginary components in the nonlinear terms. We find the stability region for a wide range of two- and three-body interaction terms with the inclusion of both gain and loss effects by using the Jacobian matrix. We investigate the stability of the system for possible different state of those cases. The stability properties of three-dimensional condensates are strongly altered by tuning the gain rate of their elastic collisions. These strong losses impose severe limitations for using Feshbach resonances. We finally sustain our semi analytical findings with the results of inclusive numerical simulations. 

\textbf{keywords:} Bose-Einstein condensates, Three-body, inelastic collision, gain/loss rate, Linear stability analysis.

\end{abstract}
\date{\today}
\maketitle

\section{Introduction}
After the first successful experimental realizations of Bose-Einstein condensates (BECs) in dilute atomic gases~\cite{Anderson1995,Bradley1995,Davis1995}, a great deal of theoretical and experimental progresses has been made in cold atom physics. The dynamical properties of BECs at low temperatures are described by the time-dependent, nonlinear, mean-field Gross-Pitaevskii equation (GPE). The qualitative nonlinearity in the condensate is induced by the inter-atomic interaction, known as $s$-wave scattering length, $a_s$. This parameter plays a key role in defining the order interaction terms considered in numerous variant forms of GPE~\cite{Sabari2014,Sabari2017a,Sabari2020a}. It is noteworthy that the theoretical studies of dynamic behaviors are mainly focused on the effect of two-body interactions. At low temperature and low density, the inter-atomic distances are much greater than the distance scale of atom-atom interactions, where two-body interaction is important and the effects of three-body interaction are negligible. But, when temperature is low with considerable high density, the magnitude of the scattering length $a_s(t)$ is much less than the thermal de Broglie wavelength. On the other hand, progress with BEC on the surface of atomic chips and atomic waveguides involves a strong compression of the BEC and an essential increase of its density. Then the problem arises of taking into account three-body interaction effects, which can start to play an important role~\cite{Sasi2025,Gammal2000,Abdullaev2001}. Recently, many works have studied the BECs with many-body interactions~\cite{Wamba2008,Josser1997,Sabari2020c,Sabari2010,Sabari2013,Sabari2015a,Sabari2015b,Sabari2018b,Sabari2019a,Sabari2020b,Sabari2021,Sabari2022a,Sabari2022b,Sabari2022c}. This interaction is interesting also for an understanding of the fundamental limits of the functioning of BEC-based devices. Nowadays, it is widely accepted that, even at a very dilute limit, two-body interactions and many-body interactions are equally important~\cite{Sabari2020c,Sabari2022a,Sabari2023,Sabari2020c,Sabari2020a,Sabari2018c,Sabari2018a}. 

At absolute zero temperature, we consider only the elastic collisions between atoms in the condensate i.e., the imaginary part of the two- and three-body interactions terms are negligible and only their real part is included ~\cite{Gammal2000,Abdullaev2001,Wamba2008,Perez1998}. But, at finite temperature ($\simeq$ nano K), we consider both the elastic and inelastic collisions between atoms in the condensate i.e., the real and imaginary parts of the two- and three-body interactions are important~\cite{Serkin2010,Sabari2016a} in the study of the condensate's dynamical properties such as stability. Note that the dynamics and stability of BECs are strongly influenced by the $a_s$. Actually, if $a_s>0$, the interactions are repulsive, the condensate is stable and its size and number then have no fundamental limit. On the other hand, for attractive interactions ($a_s<0$), the condensate is stable upto a critical value of the number of atoms due to matter wave collapses. With a supply of atoms from an external source due to elastic/inelastic collisions~\cite{Stenger1999} the condensate can grow again and thus a series of collapses can take place: this has been observed experimentally in BECs of $^7${Li} with attractive interaction~\cite{Bradley1995,Bradley1997}.

It is important to note that most of the research works on the dynamics and stability of BECs have been limited to quasi-one-dimensional (quasi-1D) cases~\cite{Zhang2007,Kengne2008}. However, even though several authors have reported the dynamics of BEC with and without trap, the stability region of multidimensional BECs in the presence of three-body interaction still deserve careful attention of researchers in this domain. Perhaps one of the reasons is that all localized solutions are usually unstable for the high-dimensional equations with constant-coefficient due to weak and strong collapses~\cite{Chao2011}. Relatively less research work on the high-dimensional equation has been carried out. Recently, the alteration of atomic interaction has been used to stabilize the high-dimensional BECs~\cite{Sabari2015c,Adhikari2001,Towers2002,Saito2003}. Thus the study of the 2D and 3D variable-coefficient nonlinear Schr\"{o}dinger equation (NLSEs) or GPEs has been recently one of the central issues in the field of BECs and nonlinear optics. For example, Abdullaev et al.\cite{Gammal2001} investigated the stability of 3D trapped BECs. Chen et al.\cite{Chen2009} studied the expansion of the self-similar light bullets. Dai et al.\cite{Dai2010a,Dai2010b,Dai2011} discussed 2D and 3D non-autonomous spatiotemporal similaritons. More recently, Wang et al.\cite{Wang2010,Wu2010} investigated 2D stable solitons and vortices for BECs with spatially modulated cubic nonlinearity and a harmonic trapping potential, respectively.

To the best of our knowledge, stability domain of high-dimensional BECs with the inclusion of real and imaginary parts of the two- and three-body interactions have not been reported so far. For instence, this work aims at investigating the stability domain of 3D BEC in the presence of both two- and three-body, elastic and inelastic collisions, whose effects around Feshbach Resonnances lead to the modification of scattering length, and further, the three-body gain/loss rates~\cite{Stenger1999}. The significance of forthcoming results lies on the understanding and monitoring of technological applications of high-dimensional BEC-based devices undergoing elastic and inelastic collisions.

The organization of the present paper is as follows. In Section II, we present a brief overview of the mean-field model from which are derived the variational equations that describe the dynamics and stability of the 3D BEC, presented in Section III. Here, we give a description of the analytical method used in this work, i.e., the projection operator method (POM) with Jacobian Matrix to find fixed points (FPs) and draw the stability domains of 3D BEC for different possible cases. Section IV reports the numerical results of the time-dependent GPE with two- and three-body interactions and their loss/gain terms through split-step Crank-Nicholson (SSCN) method for different cases. Section V summarizes our work and gives an outcome of the major findings.

\section{Nonlinear mean-field model}
The GPE can be used at low temperatures to explore the macroscopic behavior of the Bose system ~\cite{Sabari2010,Sabari2013,Sabari2015a,Sabari2015b,Sabari2018b,Sabari2019a,Sabari2020b,Sabari2021,Sabari2022a,Sabari2022b,Sabari2022c}. In the presence of two- and three-body interactions with their gain/loss terms, the time-dependent GPE can be expressed as~\cite{Gammal2000,Sabari2020c, Sabari2020a}
\begin{align}
\begin{split}
 i & \hbar\frac{\partial\Psi(\textbf{r},\tau)}{\partial \tau}= \biggr[-\frac{\hbar^2}{2m}  \nabla^2 + V(\textbf{r})+ \left(G_0+\frac{i\hbar}{2}G_1\right)*\\
& N \vert\Psi(\textbf{r},\tau)\vert^2 + \left(K_0+\frac{i\hbar}{2}K_1 \right)
 N^2 \vert \Psi(\textbf{r},\tau)\vert ^4\biggr]  \Psi(\textbf{r},\tau), \label{Jac1}
\end{split}
\end{align}
where $\nabla^2$ is the Laplacian operator, $m$ is the mass of a single bosonic atom, $N$ is the total number of atoms in the condensate,  $G_i$'s and $K_i$'s, $i = 0, 1$ are the nonlinear coefficients corresponding to the two- and three-body interactions, respectively, and $ V(\textbf{r})$ is the external trap potential. The terms $G_1$ and $K_1$ denote two-body and three-body recombination rate coefficients respectively. The positive and negative values of $G_1$ and $K_1$ correspond to gain and loss of atoms due to two- and three-body interactions, respectively. The normalization condition for the condensate wavefunction reads $\int d\textbf{r} \vert \Psi(\textbf{r}, \tau)\vert ^2 = 1$.

It is convenient to use dimensionless form of variables defined by $r = \sqrt{2}\textbf{r}/l$, $t = \tau\omega$, $l = \sqrt{\hbar/(m\omega)}$ and  $\Psi(\textbf{r},\tau)=\varphi(r,\tau)(l^3/2\sqrt{2})^{1/2}$. Then the GPE (\ref{Jac1}) in the absence of any external potential reads
\begin{align}
\begin{split}
i\frac{\partial}{\partial t}  \psi(r,t)   & = \biggr[-\frac{\partial^2}{\partial r^2} - \frac{2}{r} \frac{\partial }{\partial r}+  \left(g_0+ig_1\right)\vert \psi(r,t)\vert ^2\\
& + (\kappa_0 +i\kappa_1) \vert \psi(r,t)\vert ^4 \biggr] \psi(r,t), \label{Jac2}
\end{split}
\end{align}
where we have set $\hbar$ and $m$ to unity, then consider $g$ and $\kappa$ to be the rescaled strengths of two- and three-body interactions, respectively. The imaginary parts $g_1$ and $\kappa_1$ correspond to the rescaled two- and three-body loss or gain depending on their signs~\cite{Stenger1999}. The modified normalization condition in 3D becomes $4 \pi \int_{0}^{\infty} r^2 dr \vert \psi(r,t)\vert^2 = 1$.

\section{Variational equations}

The projection operator method (POM) has been extensively used to study complex equations like the NLSE describing the light propagation in optical fibers~\cite{TchofoDinda2001,Nakkeeran2005,Kamagate2009,Kadiroglu2009}. Here we perform such approach in the case of BECs where the governing equation is a function of $t$. Let us introduce the ansatz function as $\psi$ depending on A, R, $\beta$ and $\alpha$, which are the condensate parameters also called collective variables, namely, amplitude, width, chirp and phase respectively, dependent on $r$ and $t$. We use a generalized projection operator $P_k = \exp(i\theta)\psi_{X_k}^*$, where $\theta$ is an arbitrary phase constant and $x \in \left\{ A, R, \beta, \alpha \right\}$. By substituting the ansatz function $\psi$ in Eq.~(\ref{Jac2}), multiplying the resulting equation by $P_k$ and then integrating, we obtain the following equation:
\begin{eqnarray}
&  4\pi\int^{\infty}_{0}\Big\{ \Im\left[\psi_t\psi_{X_k}^* e^{i\theta}\right] - \Re\left[\psi_{rr}\psi_{X_k}^* e^{i\theta}\right] - \Re\left[\frac{2}{r}\psi_{r}\psi_{X_k}^* e^{i\theta}\right]\nonumber \\
& \;\;+ \left( \Re \left[g_0 \psi\psi_{X_k}^* e^{i\theta}\right]- \Im \left[g_1 \psi\psi_{X_k}^* e^{i\theta}\right]\right) \vert \psi\vert ^2 \nonumber \\
& + \left( \Re \left[\kappa_0 \psi\psi_{X_k}^* e^{i\theta}\right]- \Im \left[\kappa_1 \psi\psi_{X_k}^* e^{i\theta}\right]\right) \vert \psi\vert ^4 \Big\} r^2 dr = 0, \label{Jac3}
\end{eqnarray}%
where $\psi_{X_k}^* = \partial \psi^*/\partial X_k$, $X_k \in \{A, R, \beta, \alpha\}$, $\Im [\ldots]$ represents the imaginary part of $[\ldots]$, and $\Re [\ldots]$ corresponds to the real part of $[\ldots]$. When the phase constant $\theta$ in the above Eq.~(\ref{Jac3}) is chosen as $\pi/2$, the corresponding projection operator scheme is equivalent to the Lagrangian variational method~\cite{Nakkeeran2005}. Substituting $\theta= \pi/2$ in Eq.~(\ref{Jac3}), we obtain the equivalent Lagrangian variation as
\begin{eqnarray}
&  4\pi\int^{\infty}_{0}\Big\{ \Re\left[\psi_t\psi_{X_k}^*\right] - \Im\left[\psi_{rr}\psi_{X_k}^*\right] - \Im\left[\frac{2}{r}\psi_{r}\psi_{X_k}^*\right] \nonumber \\
& \;\;+ \left( \Im \left[g_0 \psi\psi_{X_k}^*\right]- \Re \left[g_1 \psi\psi_{X_k}^*\right]\right) \vert \psi\vert ^2 \nonumber \\
& \;\;\;\;\;\; + \left( \Im \left[\kappa_0 \psi\psi_{X_k}^*\right]- \Re \left[\kappa_1 \psi\psi_{X_k}^*\right]\right) \vert \psi\vert ^4 \Big\} r^2 dr = 0, \label{Jac4}
\end{eqnarray}
In other words Eq.~(\ref{Jac4}) is equivalent to the variations~\cite{Nakkeeran2005} of the form
\begin{eqnarray}
 \frac{d}{dt} \left( \frac{\partial  L_{eff}}{\partial \dot X_k}\right)-\frac{\partial  L_{eff}}{\partial X_k} = 0 \label{Jac4a},
\end{eqnarray}
The effective Lagrangian $L_{eff}$ is calculated by integrating the average Lagrangian density as $ L_{eff} = 4 \pi \int_0^\infty  \mathcal{L}(t)\,r^2 dr$. The average Lagrangian density of Eq.~(\ref{Jac2}) is obtained in the form
\begin{eqnarray}
\mathcal{L}(t) = &\,  \left [\frac{i}{2} \left( \psi_t \psi^* - \psi_t^* \psi \right) - \vert \nabla \psi\vert^2 \right.\nonumber \\
& \, \left.-\frac{1}{2}(g_0 + ig_1) \vert  \psi\vert^4 - \frac{1}{3}(\kappa + i\kappa_1) \vert  \psi\vert^6 \right]. \label{Jac6a}
\end{eqnarray}
When $g_1 = 0$ and $\kappa_1 = 0$, the results of Eq.~(\ref{Jac4}) and Eq.~(\ref{Jac4a}) are equivalent.

On the other hand, if $\theta= 0$ in Eq.~(\ref{Jac3}), we get the minimization of residual field which is equivalent to the bare approximation of the collective variable theory~\cite{Nakkeeran2005}
\begin{eqnarray}
&  4\pi\int^{\infty}_{0}\Big\{ \Im\left[\psi_t\psi_{X_k}^*\right] - \Re\left[\psi_{rr}\psi_{X_k}^*\right] - \Re\left[\frac{2}{r}\psi_{r}\psi_{X_k}^*\right] \nonumber \\
& \;\;+ \left( \Re \left[g_0 \psi\psi_{X_k}^*\right]- \Im \left[g_1 \psi\psi_{X_k}^*\right]\right) \vert \psi\vert ^2 \nonumber \\
& \;\;\;\;\;\; + \left( \Re \left[\kappa_0 \psi\psi_{X_k}^*\right]- \Im \left[\kappa_1 \psi\psi_{X_k}^*\right]\right) \vert \psi\vert ^4 \Big\} r^2 dr = 0, \label{Jac5}
\end{eqnarray}
In the present study, we shall use Eq.(\ref{Jac4}) which is equivalent to the Lagrangian variation. In order to obtain the governing equation of motions of the condensate parameters, the following Gaussian ansatz has been used~\cite{Kamagate2009},
\begin{eqnarray}
\psi(r,t) = A(t)\exp{\left[-\frac{r^2}{2R(t)^2}+\frac{i}{2} \beta(t)r^2+i\alpha(t) \right]}, \label{Jac6}
\end{eqnarray}
where $A(t)$, $R(t)$, $\beta(t)$ and $\alpha(t)$ are the amplitude, width, chirp and phase, respectively.
Combining Eqs.(\ref{Jac6},\ref{Jac6a} and \ref{Jac4a}) we derive the set of first order autonomous equations, equivalent to the dynamics of the system.
\begin{eqnarray}
\dot{A} & = \frac{7 g_1}{8\sqrt{2}}A(t)^3 + \frac{2 \kappa_1}{3\sqrt{3}}A(t)^5+3A(t)\beta(t), \label{Jac7a}\\
\dot{R} & = -\left[ \frac{g_1}{4\sqrt{2}}A(t)^2 + \frac{2\kappa_1}{9\sqrt{3}}A(t)^4 + 2\beta(t) \right] R(t), \label{Jac7b}\\
\dot{\beta} & = 2\beta(t)^2-\frac{2}{R(t)^4}-\frac{g_0}{2\sqrt{2}} \frac{A(t)^2 }{R(t)^2}-\frac{4 \kappa_0}{9\sqrt{3}} \frac{A(t)^4}{R(t)^2}, \label{Jac7c}\\
\dot{\alpha} & = \frac{7 g_0}{8\sqrt{2}}A(t)^2+\frac{2 \kappa_0}{3\sqrt{3}}A(t)^4+\frac{3}{R(t)^2}. \label{Jac7d}
\end{eqnarray}
The parameters of the above variational Eqs. (\ref{Jac7a}-\ref{Jac7d}) describe the dynamics of the considered BEC system. Now, using the above equations, one can exactly study the stability of the system.

\section{Fixed points and the stability domains}

From the coupled ODEs, the Jacobian matrix has been obtained to investigate the stability domain for various physical parameters. The fixed points (FPs) of the system are found by imposing the left-hand side of Eqs (\ref{Jac7a}-\ref{Jac7d}) to be zero, i.e., $\dot{X}=0$, where $X$ represents $A, R, \beta$ and $\alpha$. The threshold of existence of FPs can therefore be estimated. The stability of the FPs is determined by analysing of the eigenvalues $\lambda_j$ (with $j=1,2,3,4$) of the Jacobian matrix $M_{ij}= \partial \dot{X_i} /\partial \dot{X_j}$. The stability criterion states that if the real part of at least one of the eigenvalues is positive, then the corresponding FP is unstable. It is well known that the stable fixed points correspond to stable solutions of the GPE.

The following Matrix has then been obtained:

\begin{eqnarray}
\left(
\begin{array}{llll}
M_{11} & M_{12} & M_{13} & M_{14} \\
M_{21} & M_{22} & M_{23} & M_{24} \\
M_{31} & M_{32} & M_{33} & M_{34} \\
M_{41} & M_{42} & M_{43} & M_{44}
\end{array}
\right)
\end{eqnarray}
where
\begin{eqnarray*}
M_{11}&=&-\frac{15 A(t)^2 g_1}{4 \sqrt{2}}-\frac{20 A(t)^4 \xi _1}{3 \sqrt{3}}-\beta(t),\\
M_{12}&=&0,\\
M_{13}&=&-A(t),\\
M_{14}&=&0,\\
M_{21}&=&\frac{A(t) R(t) g_1}{\sqrt{2}}+\frac{8 A(t)^3 R(t) \xi _1}{3 \sqrt{3}},\\
M_{22}&=&\frac{A(t)^2 g_1}{2 \sqrt{2}}+\frac{2 A(t)^4 \xi _1}{3 \sqrt{3}}+2 \beta(t),\\
M_{23}&=&2 R(t),\\
M_{24}&=&0,\\
M_{31}&=&\frac{\sqrt{2} A(t) g_0}{R(t)^2}+\frac{16 A(t)^3 \xi _0}{3 \sqrt{3} R(t)^2},\\
M_{32}&=&-\frac{8}{R(t)^5}-\frac{\sqrt{2} A(t)^2 g_0}{R(t)^3}-\frac{8 A(t)^4 \xi _0}{3 \sqrt{3} R(t)^3},\\
M_{33}&=&-4 \beta(t),\\
M_{34}&=&0,\\
M_{41}&=&-\frac{5 A(t) g_0}{2 \sqrt{2}}-\frac{16 A(t)^3 \xi _0}{3 \sqrt{3}},\\
M_{42}&=&\frac{2}{R(t)^3},\\
M_{43}&=&0,\\
M_{44}&=&0.
\end{eqnarray*}

In practice, the set of equations $\dot{X}_i=0$ is solved through fourth-order Runge-Kutta algorithm by providing an initial condition for a given set of GPE parameters. When a stable fixed point is found, say $X_{i0}$, it serves as the initial condition for solving $\dot{X}_i=0$ for the neighbouring GPE parameters. We also numerically verify  our semi analytical results by using SSCN methods~\cite{Muruganandam2009}. To solve the GPE for large nonlinearity $\vert g(t) \vert$, one may start with the Thomas-Fermi approximation for the wave function obtained by setting all the derivatives in the GPE to zero \cite{Thogersen2009,fetter2009}. Alternatively, the harmonic oscillator solution is also a good starting point for small values of nonlinearity $\vert g(t) \vert$ as in this paper. The typical discretized space and time steps for solving SSCN method is 0.01 and 0.0001. Then in the course of time iteration, the coefficient of the nonlinear term is increased from 0 at each time step. Simultaneously, the initial stage of harmonic trap is also switched off slowly by changing $d(t)$ from $1$ to $0$ until the final value of nonlinearity is attained at a certain time called time $t_0$.

\subsection{Three-body gain $(\kappa_1>0)$ and loss ($\kappa_1<0$)}

Here, we have analyzed the stability of 3D BEC by tuning the three-body gain and loss ($\kappa_1$) in the absence of two-body loss/gain ($g_1$). Actually, we have shown three different domains of stability with $g_1=0$, for (i) $\kappa_1=0$, (ii) $\kappa_1>0$ and (iii) $\kappa_1<0$.

\begin{figure}[!ht]
\begin{center}
\includegraphics[width=0.49\textwidth]{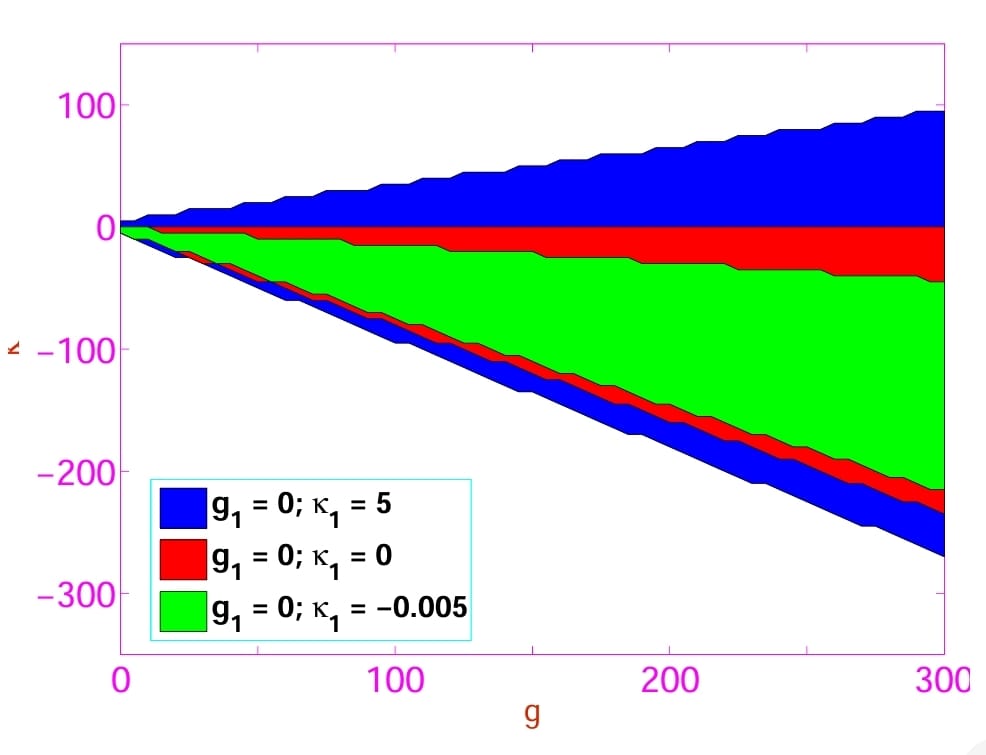}
\end{center}
\caption{Stability domain for two-body ($g$) Vs three-body ($\kappa$) with tunable $\kappa_1$, the colored/colorless domain represents the stability/instability region of BECs. (i) Red color region is for $g_1=\kappa_1=0$, (ii) blue color region is for $g_1=0, \kappa_1>0$ and (iii) green color region is for $g_1=0, \kappa_1<0$}
\label{f1}
\end{figure}

\begin{figure}[!ht]
\begin{center}
\includegraphics[width=0.49\textwidth]{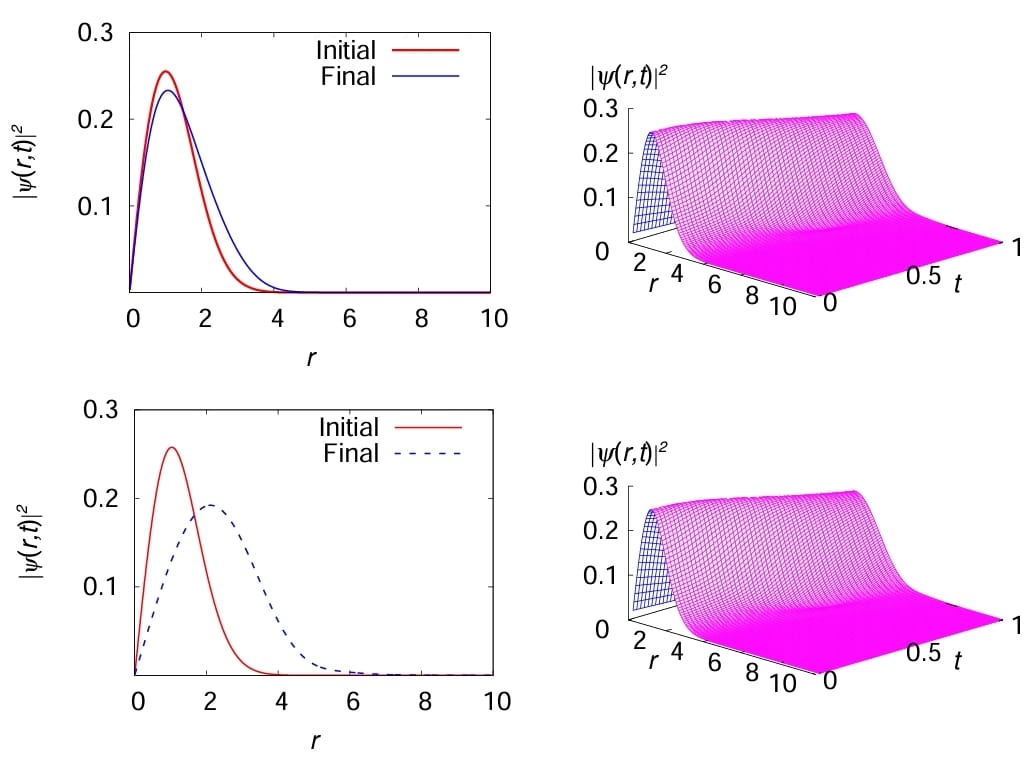}
\end{center}
\caption{Numerical plots of stability region for repulsive two-body and attractive three-body interactions ($g>0>\kappa$) with, for upper (lower) figures, both $g_1=\kappa_1=0$  ($g_1=0$ and $\kappa_1>0$).}
\label{f1a}
\end{figure}

\begin{figure}[!ht]
\begin{center}
\includegraphics[width=0.49\textwidth]{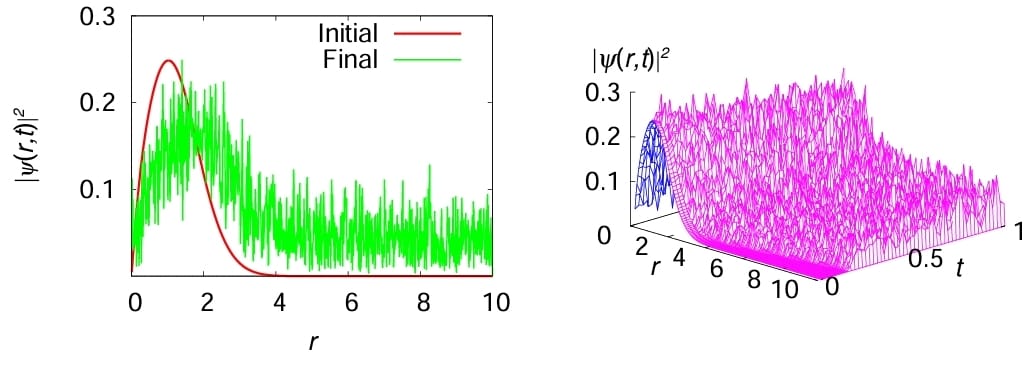}
\end{center}
\caption{Numerical plots for instability region. Here, the parameters values are chosen from the outside of the domain region.}
\label{f1b}
\end{figure}

In figure \ref{f1}, colored and colorless domains respectively represent the stability and instability regions of 3D BECs. Here, we have shown three different domains for stability of the system. The red color stability domain is for the absence of both two- and three- body loss/gain ($g_1=\kappa_1=0$). This domain is occupied in the repulsive two-body and attractive  three-body interaction regimes. In the presence of repulsive two-body interaction, the trapless system expands to infinity. Here, the occurrence of stability region is the repulsive force balanced by the three-body attraction.

If we consider the three-body gain ($\kappa_1>0$) (that is feeding the condensate from thermal clouds, system feels more repulsive effect), it gives rise to a much broader stability region of the system. But, if we include the three-body loss ($\kappa_1<0$), the stability region of the system increases. In figure \ref{f1a}, the stable density profile illustrates the dynamical stability of the 3D BEC within the domain. Here, the parameter values are taken from the middle of the domain for (i) upper rows $g>0>\kappa$ and both $g_1=\kappa_1=0$ and (ii) lower rows $g>0<\kappa$, $g_1=0$ and $\kappa_1>0$. In figure \ref{f1b}, the density profile shows the stability properties of the 3D BEC outside of the domain region. ( Roughly $g>0>>\kappa$ and both $g_1=\kappa_1=0$ ). 

\subsection{Two-body gain ($g_1>0$) and loss ($g_1<0$) with three-body gain ($\kappa_1>0$)}

\begin{figure}[!ht]
\begin{center}
\includegraphics[width=0.4\textwidth]{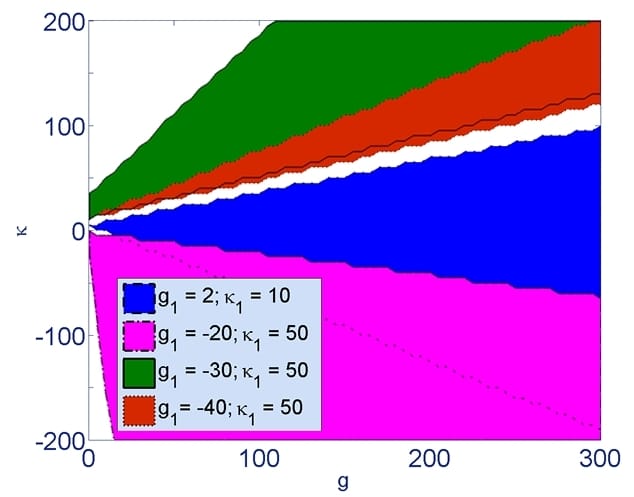}
\end{center}
\caption{Two-body ($g$) Vs three-body ($\kappa$) for two-body loss and gain ($g_1>0>g_1$) with three-body gain $\kappa_1>0$. }
\label{f2}
\end{figure}

In this subsection, we have analyzed the stability of the system by tuning the $g_1$  with the three-body gain ($\kappa_1>0$).
In figure \ref{f2}, we have shown four different domains of stability for different $g_1$ values.
With the presence of both two- and three-body gain ($g_1,\kappa_1>0$), the stability domain is equally occupied in the repulsive two-body and attractive three-body regimes.
If we tune the $g_1<0$ (from $>0$), the full domain is occupied in the repulsive two-body and attractive three-body regimes.
If we increase the two-body loss effects in the condensate ($g_1=-30$ and $-40$), the stability domain fully exist in the repulsive two- and three-body interaction regime. Here, the two-body loss effect is actualy counterbalanced by the both two- and three-body repulsive interactions.

\begin{figure}[!ht]
\begin{center}
\includegraphics[width=0.49\textwidth]{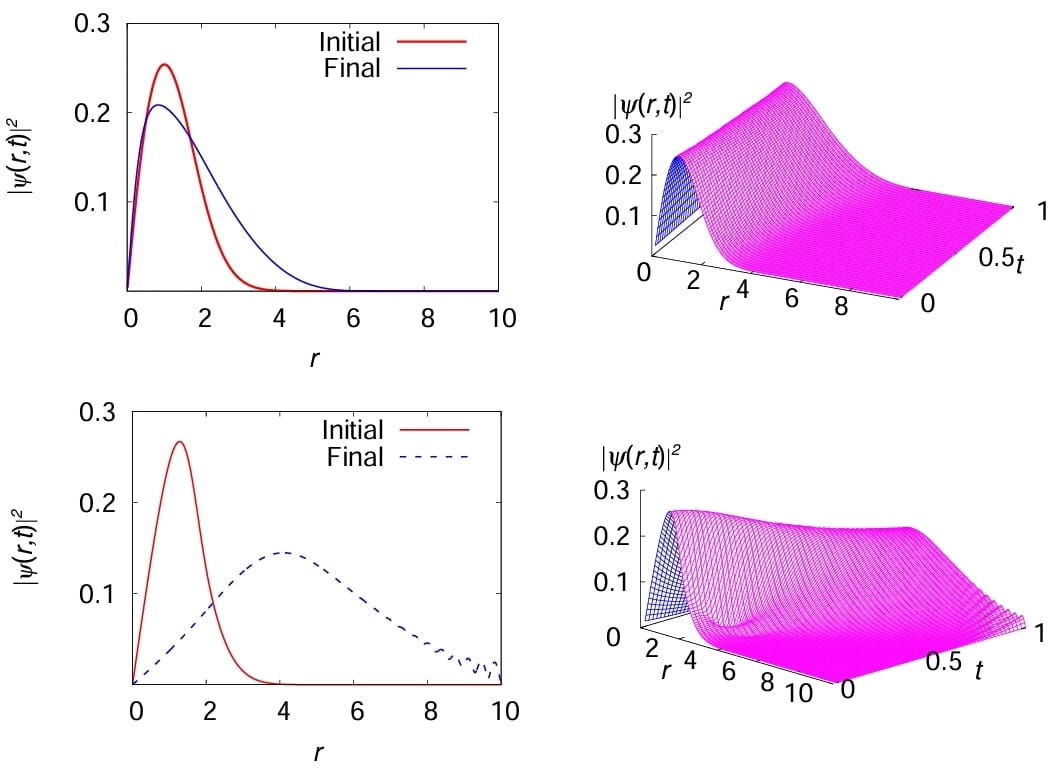}
\end{center}
\caption{Numerical plots for stability region for $g>0>\kappa$ with for (i) upper figs, $g_1>0<\kappa_1$  with their gain ($g_1>0<\kappa_1$), and (ii) lower figs, ($g>0<\kappa$) with their loss and gain ($g_1<0<\kappa_1$) (green color region in fig \ref{f2}).}
\label{f2a2}
\end{figure}

\begin{figure}[!ht]
\begin{center}
\includegraphics[width=0.49\textwidth]{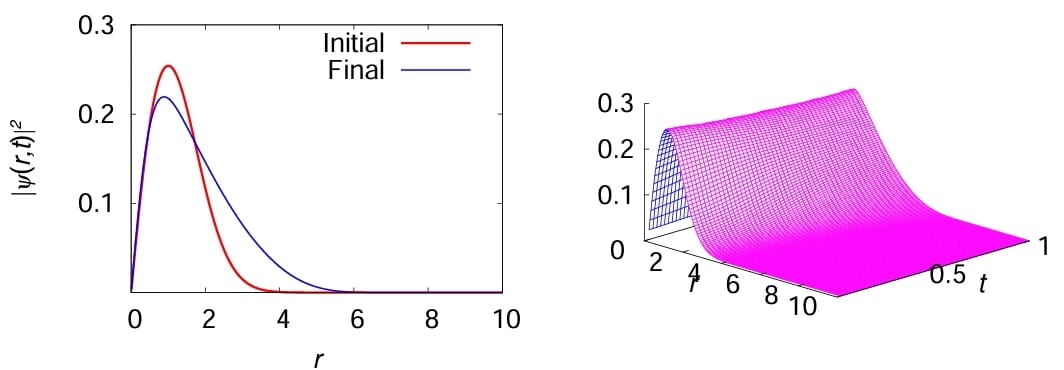}
\end{center}
\caption{Numerical plots for stability region, the values taken from the $g>0>\kappa$ and $g_1<0<\kappa_1$ region.}
\label{f2a3}
\end{figure}

Upper pannels in figure \ref{f2a2} show the density profile and space-time plot of the density for $g>0>\kappa$ with $g_1>0<\kappa_1$ and their gain ($g_1>0<\kappa_1$). Lower pannels in \ref{f2a2} show the stability of the system for both repulsive two- and three-body interactions ($g>0<\kappa$) with their loss and gain ($g_1<0<\kappa_1$), respectively. The system expands with respect to time due to the both repulsive forces, which is clearly illustrated in density plot. Although the system is stable, but for $g>0>\kappa$ with $g_1<0<\kappa_1$, it expands very slowly when compared to the previous one (figure \ref{f2a3}).

\subsection{Two-body gain and loss with three-body loss ($\kappa_1<0$)}

\begin{figure}[!ht]
\begin{center}
\includegraphics[width=0.45\textwidth]{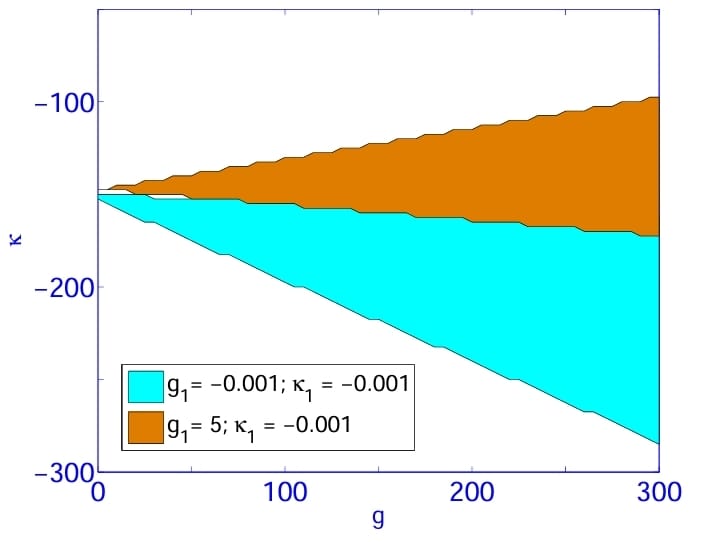}
\end{center}
\caption{Two-body Vs three-body for two-body gain and loss ($g_1>0$ and $g_1<0$) with three-body loss ($\kappa_1<0$).}
\label{f3}
\end{figure}

Lastly, we have analyzed the stability of the system with the inclusion of three-body loss $(\kappa_1<0)$ and two-body gain and loss ($g_1>0$ and $g_1<0$). In figure \ref{f3}, we have shown two domains of stability for two-body gain and loss.

\begin{figure}[!ht]
\begin{center}
\includegraphics[width=0.49\textwidth]{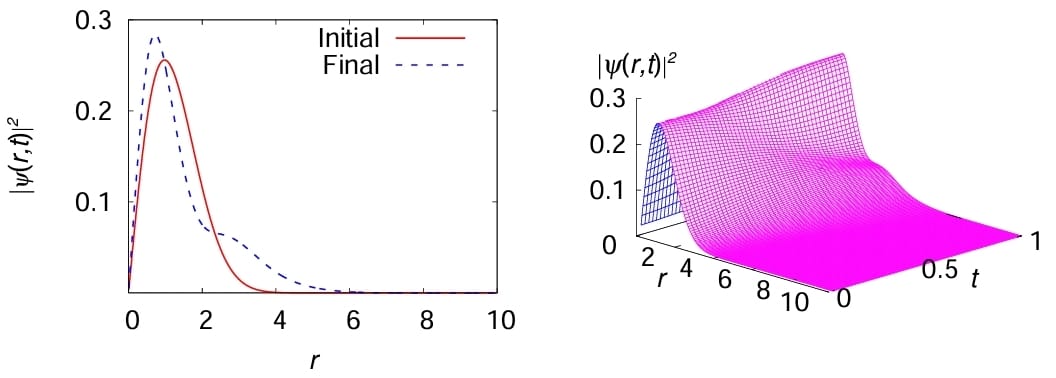}
\end{center}
\caption{Numerical plots for $g>0>\kappa$ and both $g_1=\kappa_1<0$.}
\label{f3a}
\end{figure}

Both domains are observed in the repulsive two-body and well in attractive three-body interaction regime. For the case of $g_1=\kappa_1<0$, the system is dominated by repulsive forces due to the both two- and three-body loss. So, more attractive force is needed to stabilize the system. Thus, the full domain is occupied well in the attractive three-body regime. If we tune the $g_1>0$ from $<0$, the occupation of the stability domain moves toward the repulsive three-body interaction regimes.
 
In figure \ref{f3a}, the density profile and space-time plot of the density illustrate the dynamical stability of the system within the domain region. Here, the density profile extended in the time axis due to the very high attractive three-body interaction.

In previous three case studies, it is clearly observed from the numerical plots that the analytically found stability domains are in good agreement with the numerical simulations. This is always the case, without loss of generality, mainly in repulsive two-body and attractive three-body plane.

\section{Conclusion}
Although in low dimensional BEC the stability of the system is enhanced by the inclusion of three-body interaction terms (see refs. \cite{Sabari2010,Sabari2019b,Sabari2019c}), we have questioned the stability of higher-dimensional, three-dimensional, BECs with both two- and three-body interactions under elastic and inelastic collisions. These cases were known to generate gain or loss rate of Bosons in the system. The inclusion of all above considerations at low temperatures, has enabled to describe the dynamics with a GPE containing real and imaginary parts of the cubic and quintic nonlinear terms. From the variational equations thereafter obtained, the stability domains were depicted by making good use of the Jacobian matrix method. The stability domains for possible different cases (loss or gain rates) have then been achieved, mainly within repulsive two-body and attractive three-body plane. It appears that real and imaginary parts in two- and three-body interaction terms strongly affect the stability bandwidth of the system due to external forces developed by the elastic/inelastic collisions. We have also verified our analytically found stability domains by numerical simulation through SSCN method. It always appeared that the numerical results are in good agreement with the results obtained by using semi-analytical method. Our findings are of significant importance to understand the stability of high-dimensional BECs with two- and three-body elastic/inelastic collisions.

\section*{Acknowledgements}
\noindent SS acknowledges the Funda\c c\~ao de Amparo \`a Pesquisa do Estado de S\~ao Paulo (FAPESP) [Contracts No. 2020/02185-1 and No. 2017/05660-0].


\providecommand{\newblock}{}

\end{document}